\documentclass[11pt]{article}

\usepackage[a4paper,margin=1in]{geometry}
\usepackage{authblk}

\usepackage[utf8]{inputenc}
\usepackage[T1]{fontenc}
\usepackage{lmodern}
\usepackage{microtype}
\usepackage{amsmath,amssymb}
\usepackage{graphicx}
\usepackage{booktabs}
\usepackage{xcolor}
\usepackage{enumitem}
\usepackage{url}
\usepackage[numbers,sort&compress]{natbib}
\usepackage[hidelinks]{hyperref}

\newcommand{\CD}{\mathrm{CD}_5}
\newcommand{\CDnok}{\mathrm{CD}^{\mathrm{nok}}}
\newcommand{\relies}{\textsf{relies}}
\newcommand{\departs}{\textsf{departs}}
\newcommand{\na}{\textsf{na}}

\newcommand{\figfile}[2]{%
  \IfFileExists{#1}{\includegraphics[width=#2\linewidth]{#1}}%
  {\fbox{\parbox{#2\linewidth}{\centering\vspace{3em}\textit{[figure file \texttt{\detokenize{#1}} not found]}\vspace{3em}}}}}

\title{Citations Are Late.\\
\large Reading epistemic instability from what papers believe,\\
years before the citation graph catches up}

\author[1,2]{Andrey Ustyuzhanin}
\author[3,1]{Ekaterina Trofimova}
\author[1]{Denis Zuenko}
\affil[1]{Constructor University}
\affil[2]{National University of Singapore}
\affil[3]{Constructor Labs}
\date{}

\begin{document}
\maketitle

\begin{abstract}
Paradigm shifts in science are visible in what researchers assert and contest before they are visible in the citation graph. We ask whether a cheap, content-level signal of epistemic instability, derived from the changing distribution of stated modelling beliefs in paper abstracts, can anticipate a paradigm shift earlier than the dominant citation-based disruption index ($\CD$). On the displacement of recurrent networks by Transformers in NLP, a pre-registered content signal crosses its detection threshold in 2016-Q1, whereas a real-time $\CD$ monitor cannot even observe the 2017 breakthrough until 2022-Q2, since $\CD$ needs a five-year forward-citation window: a lead of $\approx$25 quarters. The flat citation baseline is not an artifact of one index, as our $\CD$, a reference-normalised variant, and an authoritative precomputed index all sit near zero across the shift. The lead is also not a faster proxy: at equal latency the signal beats the content competitors tested, including a learned CD-from-text model and an embedding disruption measure, and the decomposition sees what a keyword cannot (keyword-blind AUC $\approx$0.9, reproduced on human labels). The lead over $\CD$ is an observability lead. Which signal carries it depends on the shift: belief adoption in NLP, contestation and applicability stress in vision. Measured against the breakthroughs themselves, the signal leads by five quarters in NLP and is contemporaneous in computer vision.
\end{abstract}

\section{Introduction}\label{sec:intro}

A scientific paradigm shifts first in the beliefs researchers state and dispute, and only later in the structure of who cites whom. The dominant quantitative tool for detecting such shifts is the consolidation--disruption index (CD). Originally devised for patent networks \citep{funk2017dynamic} and imported into science-of-science with a fixed five-year window ($\CD$) \citep{wu2019large,park2023papers}, CD reads the citation graph: it asks whether later work cites a focal paper instead of its predecessors. That window is a measurement convention rather than a claim about how fast paradigms turn over, and it makes the index structurally late: a 2017 breakthrough cannot be registered until 2022. It is also, by a growing consensus, confounded, conflating genuine disruption with citation inflation and reference-coverage artifacts \citep{park2023papers,petersen2024disruption,leibel2026multiverse}. Lateness here is constitutive of the measure, not a parameter one can tune. If we want to detect a paradigm shift as it happens, we need a different substrate.

We study a simple alternative: a content-level signal of epistemic instability, computed from the beliefs papers state in their abstracts. Our test case is a well-documented shift, the displacement of recurrent networks by Transformers in NLP \citep{bahdanau2015neural,vaswani2017attention}. From title and abstract only, we extract whether each paper assumes the incumbent paradigm, contests it, or stresses its limitations, and aggregate these into quarterly signals. We then ask a single question: does the content signal cross a pre-registered detection threshold before the citation-based index could?

The main contributions are the following.
\begin{enumerate}[leftmargin=*]
\item \textbf{A measurable, replicating lead.} A content signal detects the RNN$\to$Transformer shift in 2016-Q1, $\approx$6 years before the earliest quarter in which a $\CD$ monitor could observe it (Section~\ref{sec:nlp} and Figure~\ref{fig:cscl}). The same pipeline, unmodified, reproduces the lead on a structurally different, cross-subfield shift (hand-engineered features to deep CNNs in vision) by 14--19 quarters (Section~\ref{sec:vision} and Figure~\ref{fig:cscv}).
\item \textbf{The gap is not a baseline artifact.} The result is triangulated on both sides of the comparison. On the citation side, three independent disruption measures (our $\CD$, a reference-normalised variant, and an authoritative precomputed index from SciSciNet) all sit near zero across the shift, ruling out the ``use a better index'' response (Section~\ref{sec:nlp}). On the content side, at equal latency the signal beats the competitors tested, including the two strongest named successors to CD, none of which onset for the shift (Section~\ref{sec:bakeoff}).
\item \textbf{The decomposition is necessary and tractable.} A component decomposition recovers the shift where a keyword counter is blind, recovers the incumbent axis without an oracle, and forms a low-dimensional, drift-stationary sufficient statistic (Section~\ref{sec:decomp}).
\item \textbf{Human validation confirms the construct and its necessity result.} A first human-annotated slice confirms the paradigm-reliance construct is human-recoverable on in-scope items ($\kappa\approx0.83$) and free of fabricated warrants (100\% verbatim), and, critically, the keyword-blind advantage from Contribution 3 reproduces on human labels, converting that finding from an LLM-panel result to a human-validated one (Section~\ref{sec:human}).
\end{enumerate}

\section{Related Work}\label{sec:related}

\paragraph{Citation-based disruption indices.} The consolidation--disruption (CD) index \citep{funk2017dynamic} and its variant for science \citep{wu2019large} classify a paper as disruptive when subsequent work cites it in place of its intellectual predecessors. The index has driven high-profile findings, notably a claimed secular decline in disruptiveness \citep{park2023papers}. However, a substantial methodological literature now documents that CD is sensitive to citation inflation, reference counts, and database coverage, and that a large ``multiverse'' of implementation choices changes the result \citep{petersen2024disruption,leibel2026multiverse}. We take this as settled background rather than a target: our comparison uses $\CD$ as the incumbent baseline and does not attempt to relitigate its flaws. Two properties matter for us and are not in dispute: CD requires a multi-year forward-citation window (it is structurally lagged), and it is content-blind (it reads only the citation graph).

\paragraph{Content- and text-based signals of change.} A separate line of work reads the text itself: atypical recombination of prior work as a novelty signal \citep{uzzi2013atypical}, citation-intent and argumentative framing \citep{cohan2019structural}, and language-model surprise. These are computable at publication time, making them candidates for early detection; what has been missing is a direct, pre-registered head-to-head against the citation baseline on documented shifts, with a lead-time estimate. Our epistemic-instability signal, i.e., the share of papers that adopt versus contest a modelling paradigm, sits in this family and is designed to fire on the belief collapse that characterises a Kuhnian shift \citep{kuhn1970structure}.

\paragraph{Diachronic text change and changepoint detection.} A large literature detects change directly from text: diachronic word embeddings and semantic-shift detection \citep{hamilton2016diachronic,kutuzov2018diachronic}, dynamic topic models \citep{blei2006dynamic}, and time-series changepoint detection \citep{aminikhanghahi2017survey}. These track distributional drift in words or topics, which is powerful but diffuse: a paradigm shift and a mere vocabulary fashion both register as drift, and thresholding a smooth drift on a gradual transition is hard. Our signal instead targets a specific, interpretable proposition: reliance on the incumbent paradigm, judged by the role of a method rather than the presence of a term. This separates genuine departure from vocabulary drift, since a keyword or embedding-drift signal counts an ``LSTM with attention'' as change while ours does not, and yields a low-dimensional, stationary monitor rather than a high-dimensional, drifting one (Section~\ref{sec:decomp}). We view these as complementary and, on the gradual shifts we study, expect belief-level decomposition to be sharper than a topic- or semantic-drift signal. A further direction is anticipatory statistics from complex-systems theory, where rising variance and autocorrelation precede critical transitions \citep{scheffer2009early}; applying them to a content signal is a natural next step and a limitation of the present fixed-threshold detector.

\section{Method}\label{sec:method}

Our design is pre-registered in an analysis plan committed before the detection step; all thresholds, windows, and the control topic were fixed in advance. We take arXiv metadata (title and abstract only) for a field and window bracketing a documented shift, and apply a cheap keyword pre-filter to the sub-topic where the shift occurred (reported filter precision 0.72 on a 50-paper hand check for NLP); no full text or external knowledge is used downstream. From each abstract, a low-cost language model (\texttt{gpt-4o-mini}, temperature 0) extracts four fields: the paper's architecture family; whether its method assumes the incumbent paradigm; whether it contests that paradigm (explicitly or implicitly); and whether it reports applicability stress (limitations of the incumbent). To bound hindsight, the prompt forbids outside knowledge and requires a verbatim evidence span for the judgement fields, and a second model of a different family (\texttt{gemini-2.5-flash}) re-extracts an independent pass for validation.

Over the papers whose architecture is applicable, we aggregate these judgements into four quarterly signals: belief adoption $A(t)$ (share assuming the incumbent), contestation $C(t)$, applicability stress $X(t)$, and a Bayesian surprise $S(t)=\mathrm{KL}(P_t\,\|\,P_{t-1})$ over the architecture-family distribution (Dirichlet-smoothed); all signals are 2-quarter trailing-smoothed. A signal is detected at the first quarter after a fixed baseline window where its smoothed value exceeds baseline $\mu+2\sigma$ for two consecutive quarters. The citation baseline's observability wall is the earliest calendar quarter a real-time $\CD$ monitor could compute the index for the flashpoint cohort, i.e., the milestone quarter plus five years, and the lead-time is $\Delta=T_{\mathrm{wall}}-T_{\mathrm{content}}$, reported per signal and positive if content leads. For the head-to-head we compute $\CD$ \citep{funk2017dynamic} from OpenAlex for the most-cited papers per year via exact facet-count queries, and additionally report a reference-normalised variant (dropping the swamping predecessor-citation term) and an authoritative precomputed disruption index from SciSciNet \citep{lin2023sciscinet}. The pipeline is open (Appendix~\ref{app:repro}); models are version-pinned and total LLM cost is under \$11 across both case studies.

\section{Case 1: Transformers displacing recurrence in NLP}\label{sec:nlp}

We evaluate the pipeline on arXiv cs.CL for the period 2013 to 2022, comprising 39,116 papers, of which 23,567 form the sequence-modelling treatment set and 7,367 an architecture-neutral control. The underlying shift is unambiguous: the share of recurrent models peaks at 43\% in 2016 and falls to 7\% by 2022, while the Transformer share rises from 0\% to 47\%.

\begin{figure}[t]
\centering
\figfile{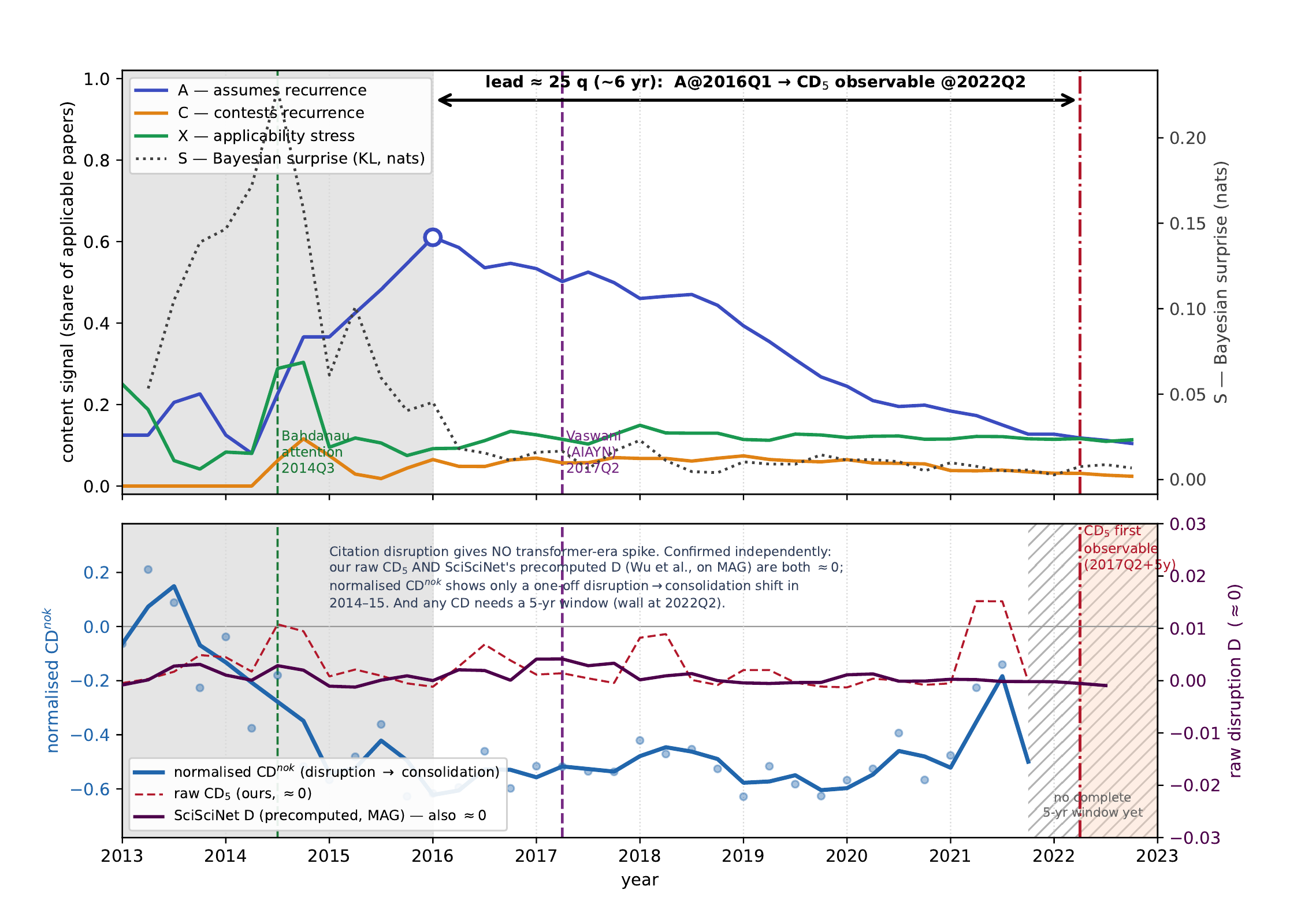}{0.95}
\caption{\textbf{cs.CL.} Top: content signals; belief adoption $A$ (blue) is detected at 2016-Q1 (circle) and the arrow marks the 25-quarter lead to the $\CD$ observability wall (2022-Q2). Bottom: three citation-disruption measures (our $\CD$, normalised $\CDnok$, and SciSciNet's precomputed $D$) are all $\approx0$ across the shift.}
\label{fig:cscl}
\end{figure}

\paragraph{The content signals yield a measurable lead over citation-based detection (P1).} Figure~\ref{fig:cscl} (top) shows that belief adoption $A$ rises to a peak of 0.61 in 2016-Q1 and then collapses; under the pre-registered decision rule, $A$ is detected at 2016-Q1. A real-time $\CD$ monitor, by contrast, cannot observe the 2017-Q2 flashpoint \citep{vaswani2017attention} until 2022-Q2, giving a lead of 25 quarters, or roughly six years. The advantage arises from an observability lag rather than from a difference in effect size: throughout that interval the citation index is unavailable, not merely small. The onset is robust to the detection parameters (Appendix~\ref{app:sens}).

\paragraph{The lead is robust to the choice of index (P1b).} Figure~\ref{fig:cscl} (bottom) plots three disruption measures on a common axis: our $\CD$, a reference-normalised $\CDnok$, and the precomputed $D$ of SciSciNet \citep{lin2023sciscinet}, evaluated on 3,967 corpus papers. All three remain close to zero and exhibit no Transformer-era spike; even landmark papers receive small and inconsistent scores (Bahdanau $+0.10$, BERT $-0.06$), and Vaswani is unmeasurable in OpenAlex owing to record fragmentation (Appendix~\ref{app:landmark}). The lead is therefore not an artifact of a poorly chosen index. Specificity is likewise supported (P2): on the matched control topic no content signal crosses threshold, indicating that the lead reflects a genuine architectural shift rather than field-wide drift. Extraction reflects contemporaneous evidence rather than hindsight (P3): a verbatim-warrant audit finds no unsupported positive labels (0 of 38; Table~\ref{tab:extraction}), which bounds post-2017 leakage without eliminating it.

\begin{table}[t]
\centering
\caption{Extraction validation. $^\dagger$Raw multiclass inter-model agreement is depressed by a \texttt{no}/\texttt{not-applicable} labelling-convention boundary that no signal uses; the signal-relevant distinctions agree at 0.85--0.93. Analyst gold was not repeated for the cs.CV replication.}
\label{tab:extraction}
\begin{tabular}{lcc}
\toprule
Check & cs.CL & cs.CV\\
\midrule
Signal-relevant agreement (analyst vs.\ model) & 0.85--0.93 & ---\\
Inter-model agreement (pass A vs.\ pass B) & 0.51--0.69$^\dagger$ & 0.60--0.80\\
Leakage audit (unsupported positive labels) & 0/38 & 0/41\\
\bottomrule
\end{tabular}
\end{table}

\paragraph{The lead is carried by a single signal (P4).} Belief adoption $A$ accounts for it, whereas contestation $C$ and stress $X$ are directionally correct but remain sub-threshold. A naive Bayesian-surprise statistic $S$ fails under the pre-registered rules: it is noise-dominated when quarterly paper counts are small (2013 and 2014) and near-silent during the gradual shift of 2018 to 2020. We report this negative result because it identifies the concrete measurement problem that a principled belief-change statistic must solve.

\section{Is it more than a faster proxy? Equal-latency content baselines}\label{sec:bakeoff}

The lead in Section~\ref{sec:nlp} could be dismissed if the content signal were merely a faster proxy for CD or a dressed-up keyword counter. We therefore ran every competitor on the \emph{same} pre-registered quarterly grid at equal latency ($t=0$, no forward-citation window), scored by the same onset rule and by the same shift-vs-(vocabulary-matched-)control AUC as our signal (Appendix~\ref{app:control}). The competitors span the weak-to-strong range named in the literature: challenger-term prevalence (N1), rolling $n$-gram surprise (B1), embedding surprise (B5), reference atypicality (B2, \citet{uzzi2013atypical}), and, crucially, the two strongest named content predictors, namely a learning-based CD-from-text model \citep{he2024predicting} and an embedding-based disruption measure \citep{kim2026uncovering}. Following the same reading rule as elsewhere, a tie or loss would be reported plainly.

\begin{table}[t]
\centering
\caption{Equal-latency bake-off (cs.CL). Onset is scored by the same pre-registered rule as the belief signal; the AUC column is shift versus a vocabulary-matched control, and values below 0.5 are anti-informative, meaning the shift locus scores lower than the control. B2 was computed for cs.CV; its cs.CL run is a separate pull.}
\label{tab:bakeoff}
\small
\begin{tabular}{llcc}
\toprule
signal (equal latency, $t=0$) & family & onset & shift-vs-control AUC\\
\midrule
A: belief adoption (ours) & content decomposition & 2016-Q1 & 0.99\\
N1: challenger-term prevalence & keyword & n/a & 0.00\\
B1: rolling $n$-gram surprise & lexical surprise & n/a & $\approx0.0$\\
B5: embedding surprise & semantic surprise & n/a & 0.02\\
B2: reference atypicality (cs.CV) & bibliometric novelty & n/a & 0.29\\
B6: SPS-BERT (CD-from-text) & learned CD successor & n/a & 0.01\\
B7: embedding-disruptiveness & embedding CD successor & n/a & 0.38\\
\bottomrule
\end{tabular}
\end{table}

Table~\ref{tab:bakeoff} is unambiguous: no equal-latency baseline onsets for the shift, and none separates it from the matched control, whereas the belief signal does both. In particular, neither of the two recent successors to CD helps. The learned CD-from-text model reproduces CD's flatness, which is expected given that it is trained to emulate a flat target, and the embedding disruption measure never rises either. The distinguishing property of the content signal is therefore content grounding rather than raw speed, which clears the objection that the signal is a faster CD or keyword proxy. Baseline implementations are given in Appendix~\ref{app:baselines}.

One asymmetry must be stated, since discrimination and lead-time rest on different properties of the corpus. The belief AUC is computed on the full corpus, whereas the citation-based successors (B6, B7) are defined only on the top-cited focal papers, as citation disruption is undefined for uncited work. Controlling for this, we recomputed the belief signal on the same focal papers used by B7: it still separates the shift from a matched control ($A$ 0.96, $C$ 0.88, $X$ 0.91) against B7's 0.38, so the discrimination advantage is not an artifact of the wider population. The lead-time additionally leverages what makes a content signal useful in the first place, namely that it is computable densely, from the abstract alone, with no citation lag. It is therefore available years before any citation measure can even be evaluated on the sparse, five-year-lagged cited subset.

\section{The decomposition sees what a keyword cannot}\label{sec:decomp}

A sharper objection than the faster-proxy one is that the semantic decomposition is superfluous, that is, that the signal reduces to whether a paper uses the challenger vocabulary. We rebut this in two ways, and then show that the decomposition additionally turns shift detection into a small, drift-robust monitoring problem.

The first is that the incumbent axis is recoverable without an oracle. We clustered the open assumptions (Task B) across the corpus without telling the model the incumbent belief. The recurrence-necessity axis re-emerges as a coherent cluster, with 78\% of its statements mentioning recurrence, RNNs or LSTMs, and a discovered-axis reliance curve correlates moderately with the hand-specified one ($r\approx0.45$). The axis is thus present in the data without being injected by construction, which defuses that concern, although it is neither dominant nor the most prominent discovered assumption (size-rank 11/16): the open assumptions are dominated by generic training, language-modelling and evaluation themes.

\begin{figure}[t]
\centering
\figfile{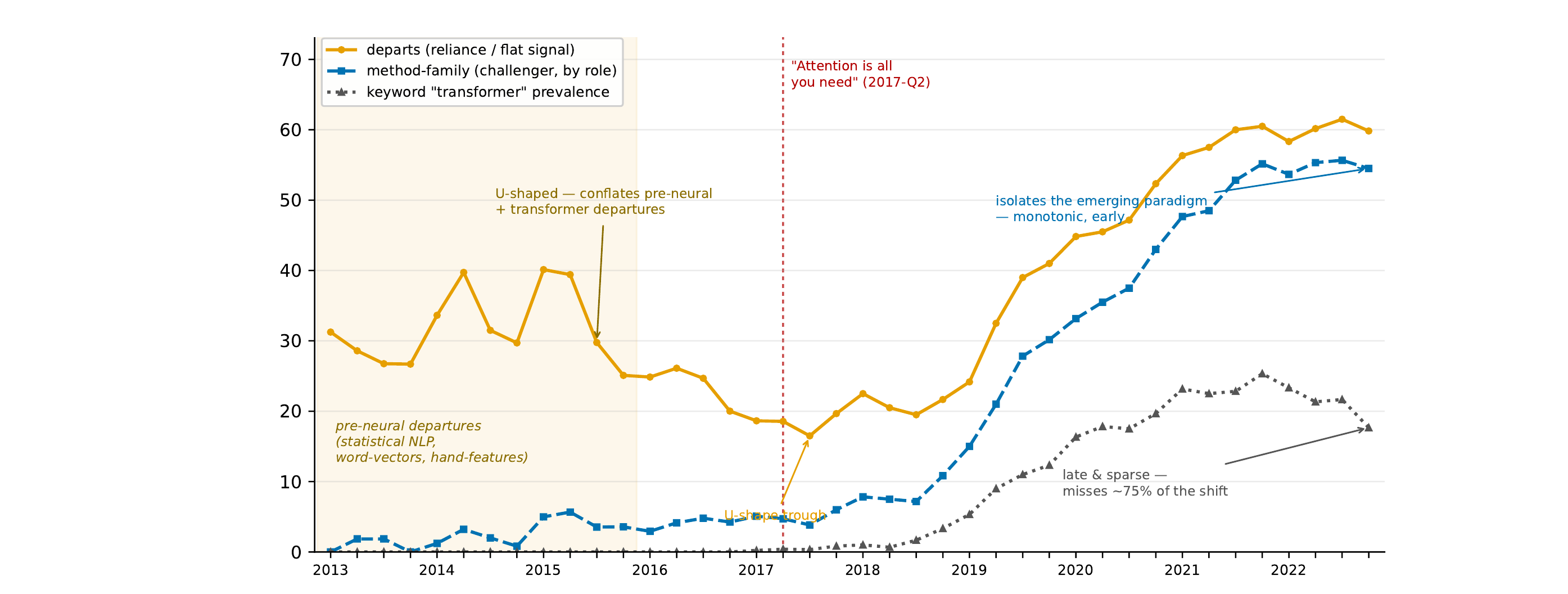}{0.95}
\caption{Which departure? cs.CL, 2013 to 2022. The keyword ``transformer'' is late and sparse; the raw \departs\ share is U-shaped, its early hump being pre-neural non-recurrent work (shaded); the decomposed method-family component is monotonic, isolates the emerging paradigm, and leads the keyword.}
\label{fig:which}
\end{figure}

The second is that the decomposition sees what a keyword counter cannot. The decisive test restricts attention to challenger-term papers, where the keyword is constant and hence uninformative (AUC $=0.5$ by construction). There, a method-family component, read by the role of the technique rather than the presence of a term, recovers the relies/departs distinction at AUC $\approx0.92$ on model labels and 0.79 on the human-adjudicated gold (Section~\ref{sec:human}). The keyword is also sparse: the literal term appears in only $\approx25\%$ of departing papers, missing three-quarters of the shift. Finally, the decomposition resolves a confound that the flat signal cannot (Figure~\ref{fig:which}). The raw departs share is U-shaped, because its early hump is pre-neural, non-recurrent work (statistical NLP, word-vectors) that also departs from recurrence, whereas the method-family component is monotonic and isolates the emerging paradigm. On a non-circular control the methods-aware component (0.83) beats the flat reliance signal (0.64), with positive incremental value after partialling out both the keyword and the flat signal ($\Delta\mathrm{AUC}\approx+0.04$).

Beyond discrimination, the decomposition has a second motivation: the resulting monitor is low-dimensional and stationary, whereas a keyword basis is neither. The keyword basis for cs.CL comprises $|V|\approx6$k unigrams, or roughly 20k with bigrams, against $\approx35$ component coordinates (reliance, method-family, assumption clusters). The decisive difference, however, is stationarity rather than size. Trained on papers up to 2017 and tested on papers from 2019 onwards, a keyword model's AUC falls from 0.91 to 0.71 and its top predictive vocabulary is $\approx97\%$ replaced (top-15 Jaccard 0.03: early ``convolutional, gram, vector'' against late ``bert, transformer, pretrained''). The component basis is instead a fixed rule. Because the shift-marking term does not exist before the shift, a keyword monitor cannot be pre-registered at all, whereas \relies/\departs\ is definable before Transformers exist.

\section{Human validation}\label{sec:human}

Every result so far is produced and cross-checked by language models. To establish that the paradigm-reliance construct is not a closed model loop but a real, human-recoverable signal, we ran a human-annotation study on a slice of the cs.CL corpus. Two independent assessors labelled a blinded, stratified slice of $\approx$100 cs.CL abstracts (the challenger-term-enriched $\kappa$-set), seeing only title and abstract, and never the model labels, the topic role, or each other's answers. They used a pre-registered codebook in which Task~A assigns the paradigm-reliance label $\{\relies,\departs,\na\}$, that is, whether the method models the sequence with recurrence as its backbone, without it, or is off-scope, together with a verbatim warrant span, and Task~B lists open assumptions without being told the incumbent axis. A short calibration round preceded independent labelling. The codebook stresses judging by the \emph{role} of the technique rather than the presence of a term, so that an ``LSTM with attention'' still counts as \relies. The codebook, both assessors' labels, the adjudication, and the resulting gold set are released \citep{ustyuzhanin2026seismic}.

Table~\ref{tab:human} summarises the study, which yields four results. On reliability, the assessors agree strongly on the substantive \relies-vs-\departs\ judgment (Cohen's $\kappa\approx0.83$ \citep{cohen1960coefficient}, with 92\% raw agreement among papers both consider in scope). The lower overall $\kappa$ of 0.55 is driven almost entirely by the off-scope (\na) boundary, that is, by whether an applied paper counts as doing sequence modelling, rather than by the reliance judgment itself; we sharpen that scope rule in the codebook and treat it as a measurement boundary rather than a construct failure. On fabrication, all warrant spans cited by the assessors are verbatim substrings of the abstract (Table~\ref{tab:human}), the human counterpart to the zero-fabrication LLM audit. On model accuracy, we resolved all 29 inter-assessor disagreements into a single gold label via a second, independent adjudication pass; on the 13 hardest cases, where the two passes still diverged (they agreed on 16/29), we tie-broke by the pre-registered codebook, judging by the backbone's role rather than the presence of a term. Against the resulting 97-item gold (departs 37, relies 31, na 29) the LLM-panel majority is correct in 74\% of cases overall and 88\% on in-scope (\relies/\departs) papers, with the residual gap lying at the \na\ boundary rather than in the reliance judgment. Finally, the necessity result reproduces on human labels. The pilot's central methodological claim, that a component decomposition detects the shift where a keyword counter is blind (Section~\ref{sec:decomp}), holds when the ground truth is human: restricted to challenger-term papers, where a keyword is uninformative by construction, a method-family component recovers the human \relies/\departs\ distinction at AUC $=0.79$ ($n=56$) against 0.5 for the keyword. This converts the finding from an LLM-panel result to a human-validated one.

\begin{table}[t]
\centering
\caption{Human-validation study on the cs.CL slice (two independent assessors, $\approx$100 abstracts). The paradigm-reliance construct is human-reliable and free of fabricated warrants, and the decomposition's advantage over a keyword counter reproduces on human labels.}
\label{tab:human}
\small
\begin{tabular}{lll}
\toprule
Measure & Value & Reference\\
\midrule
Inter-assessor $\kappa$, \relies\ vs \departs\ (in-scope) & $\approx0.83$ & gate $\ge0.6$\\
Overall $\kappa$ (incl.\ the \na/off-scope boundary) & $\approx0.55$ & scope boundary\\
Warrant validity (verbatim in abstract) & 100\% (127/127) & vs LLM audit\\
Adjudicated gold (2nd pass + codebook tie-break) & 97 items (16/29 pass-agree) & n/a\\
LLM panel vs.\ adjudicated gold & 74\% all / 88\% in-scope & n/a\\
Keyword-blind necessity on human gold (AUC) & 0.79 ($n=56$) & keyword 0.50\\
\bottomrule
\end{tabular}
\end{table}

This is a first validation slice of 97 NLP abstracts labelled by two assessors and adjudicated into a gold set, and the adjudication is itself imperfect: a second independent pass agreed with the first on only 16/29 of the pre-selected hardest disagreements, so the gold on the na boundary is codebook-defined rather than consensus-defined. What the annotation establishes is the load-bearing construct, namely paradigm reliance, warrant honesty, and the necessity of decomposition, together with a model-versus-gold accuracy of 88\% in scope; the remaining pipeline numbers are model-derived (Section~\ref{sec:limits}).

\section{Case 2: Deep CNNs displacing hand-crafted features in vision}\label{sec:vision}

To test whether the lead-time result is specific to NLP, we ran the same pipeline on a structurally different, cross-field shift: hand-engineered image features (SIFT, HOG, bag-of-visual-words, deformable parts) giving way to deep convolutional networks after AlexNet \citep{krizhevsky2012imagenet}. The corpus is arXiv cs.CV for the period 2010 to 2019, comprising 35,265 papers, of which 19,243 form the treatment set. The shift is again unambiguous: the hand-crafted share falls from 45\% in 2010 to 1\% in 2019 while the deep-CNN share rises from 0\% to roughly 79\%.

\begin{figure}[t]
\centering
\figfile{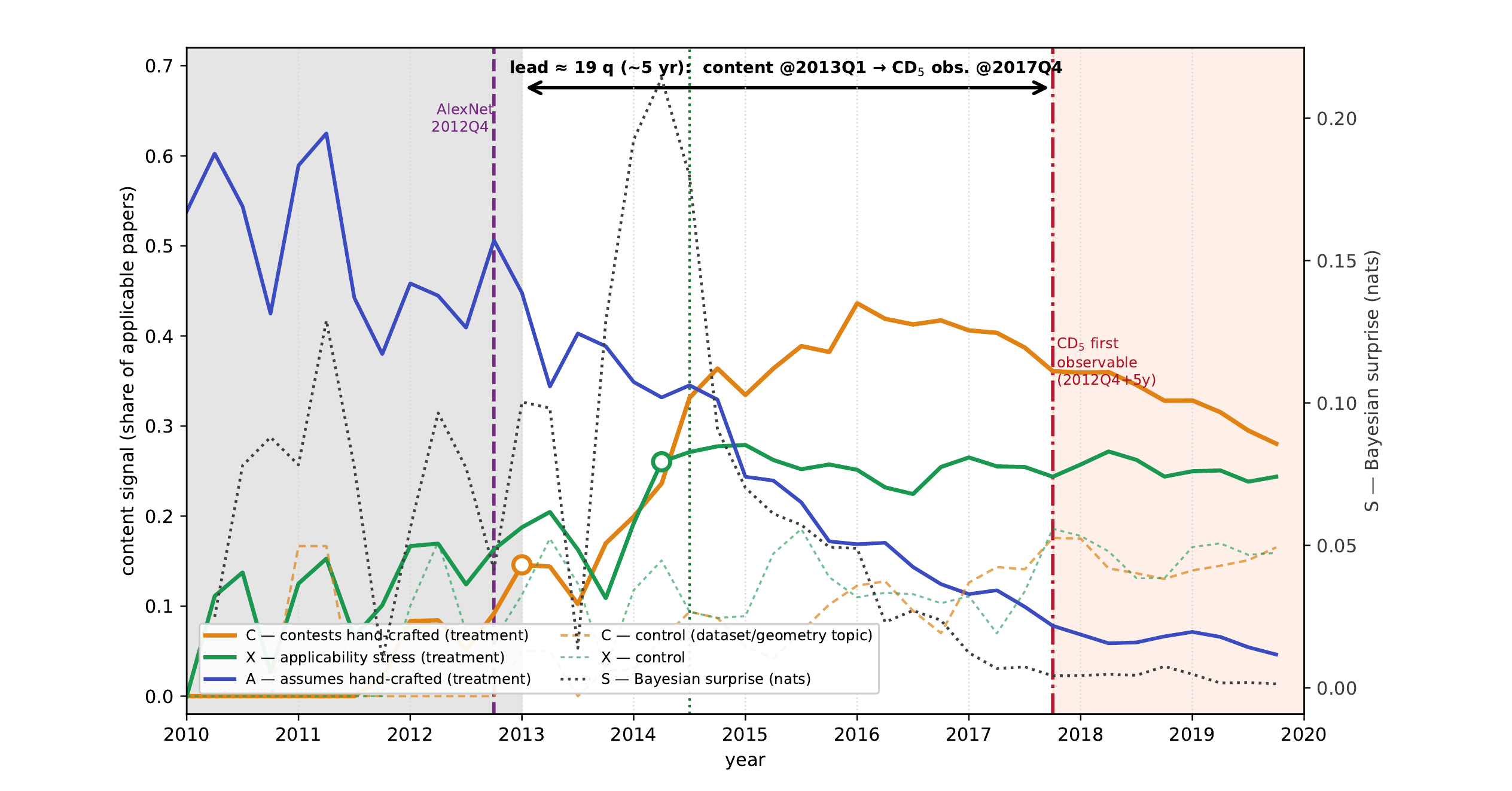}{0.95}
\caption{\textbf{cs.CV replication.} Content signals (treatment, solid) cross threshold in 2013--2014, leading the AlexNet-cohort $\CD$ observability wall (2017-Q4) by 14--19 quarters; the control (dashed) crosses much later and weaker.}
\label{fig:cscv}
\end{figure}

\paragraph{The lead reproduces (P5).} Figure~\ref{fig:cscv} shows content signals crossing threshold years before the AlexNet-cohort $\CD$ observability wall at 2017-Q4: contestation $C$ at 2013-Q1 (a lead of 19 quarters), Bayesian surprise $S$ at 2014-Q1 (15 quarters), and applicability stress $X$ at 2014-Q2 (14 quarters). The carrier differs from NLP. Here belief adoption $A$ tracks the outgoing paradigm, that is, the assumption that hand-crafted features are required, and therefore declines, while contestation and stress lead. Notably $S$ fires in this domain, where the shift is more abrupt than the gradual NLP one, which sharpens the P4 negative: the failure of naive surprise is specific to a gradual shift rather than a property of surprise in principle. Cross-field, the content-signal family thus reproduces the lead even as the individual carrier changes with the geometry of the shift.

One caveat on specificity must be stated. Unlike NLP, the vision control is not a clean null: deep learning was so pervasive that even dataset and classical-geometry papers eventually contest hand-crafted features, with control $C$ detecting only in 2017-Q4 and $X$ in 2015-Q2. The treatment nonetheless leads the control by roughly 19 quarters and at three to four times the magnitude, so the shift is still clearly distinguished from drift; a cleaner vision control, drawn from a neural sub-field that did not change its feature paradigm, is left to future work. Extraction validation for this case rests on cross-model agreement alone: inter-model agreement is 0.60 to 0.80 and the audit finds no unsupported labels (0 of 41; Table~\ref{tab:extraction}).

\section{Limitations}\label{sec:limits}

Both shifts were selected because they are documented, which is selection on the outcome, and both live on arXiv within computer science, so ``cross-field'' here means cross-subfield; a prospective test on an unresolved shift in a distant discipline would settle generality. The verbatim-evidence audit bounds hindsight without eliminating it, since a model trained after the shift may still weight in-abstract cues with foresight it did not have; a leakage-free protocol (time-stamped checkpoints, post-cutoff-only evaluation) is the clear next step. The carrier signal is domain-dependent and a naive Bayesian surprise fails on gradual shifts, which motivates a sample-size-robust belief-change statistic paired with anticipatory early-warning statistics \citep{scheffer2009early}. Detection uses a fixed $\mu+2\sigma$ rule with no interval on the lead-time and $n=2$ shifts, so quarter-level onset precision should not be over-read. Finally, extraction is abstract-only and downstream of a keyword pre-filter of precision 0.72; the vision control is contaminated rather than clean; and human validation is a first slice of two assessors and a 97-abstract NLP gold (Section~\ref{sec:human}), with the vision replication validated only by cross-model agreement. None of this touches the central negative result: across three disruption indices and two learned successors, no citation-based quantity moves when the paradigm does. The positive result is offered as a pilot on two cases.

\section{Conclusion}\label{sec:conclusion}

Reading beliefs rather than citations makes a paradigm shift observable at publication time, whereas citation-based disruption indices cannot report for years. The lead is an \emph{observability} lead and decomposes into two quantities. Most of it is the definitional lag of $\CD$: no five-year index can report on a 2017 event before 2022, whatever signal it is compared against. The smaller and more demanding quantity is the lead over the breakthrough itself, and here the two cases differ. In NLP the belief-adoption signal turns five quarters before the 2017-Q2 flashpoint, so the detection is anticipatory; in vision the earliest content signal fires one quarter \emph{after} AlexNet, so the detection is contemporaneous and the lead over $\CD$ is entirely the index's lag. The defensible claim is therefore that content-level monitoring carries zero forward-citation latency while citation-based disruption is structurally unavailable for years. Establishing anticipation as the rule rather than the NLP case requires more shifts. We release the annotation construct, the human and LLM labels, and both case studies (Appendix~\ref{app:repro}) to invite application to other fields.

\bibliographystyle{plainnat}
\bibliography{references}

\appendix

\section{Reproducibility}\label{app:repro}

The annotation construct and its revision history, the sampled papers with their arXiv identifiers, the LLM panel labels, both assessors' annotations, the adjudication, the resulting human gold set, and the annotation analysis scripts are released \citep{ustyuzhanin2026seismic}. 
The corpus is reconstructible from arXiv metadata using the released sampling description. The headline figures regenerate from committed tables with a single command; models are version-pinned and total LLM cost is under \$11 across both case studies. Citation records were resolved against OpenAlex (with a title fallback raising the match rate to 97.4\%) and cross-checked against Semantic Scholar and the SciSciNet data lake \citep{lin2023sciscinet}.

\section{Detection sensitivity}\label{app:sens}

To check that the onset is not an artifact of a single tuning, we swept the four pre-registered knobs---baseline window (three choices: 2013--2015, 2013--2014, 2013-Q1--2015-Q2), threshold $k\in\{1.5,2.0,2.5,3.0\}\sigma$, consecutive-quarter requirement $\in\{1,2,3\}$, and smoothing width $\in\{1,2,3\}$ (108 combinations). The belief-adoption signal $A$ crosses its threshold in 2015-Q1--2016-Q1 in every combination that detects at all (74\% of the grid; the pre-registered default $k=2$, consec$=2$ gives 2016-Q1). Crucially the onset only moves earlier under looser baselines and never past 2016-Q1, so the $\approx$6-year lead over the CD observability wall (2022-Q2) is intact across the grid. Non-detections occur only at the strictest settings ($k\ge2.5$ with three consecutive quarters, or no smoothing), where the adoption peak sits near $2\sigma$ above baseline; the pre-registered default is comfortably inside the detecting region. The lead-time result is thus robust to the detection parameters.

\paragraph{Uncertainty on the quarterly estimates.} Each quarterly share carries a Wilson 95\% confidence interval; the median quarter has $n\approx250$ applicable papers (range 7--1071), so the series is well-determined rather than quarter-to-quarter noise. At the onset the adoption share sits far above baseline---2016-Q1: $A=0.62$, $n=97$, CI $[0.52,0.71]$ against a baseline mean of $\approx0.20$---and the neighbouring quarters agree (2015-Q4 $[0.50,0.70]$, 2016-Q2 $[0.48,0.62]$). The residual uncertainty is in the exact onset quarter, consistent with the 2015-Q1--2016-Q1 spread of the sweep above, not in whether a large, well-separated rise occurs.

\section{Landmark-paper disruption scores}\label{app:landmark}

\begin{figure}[t]
\centering
\figfile{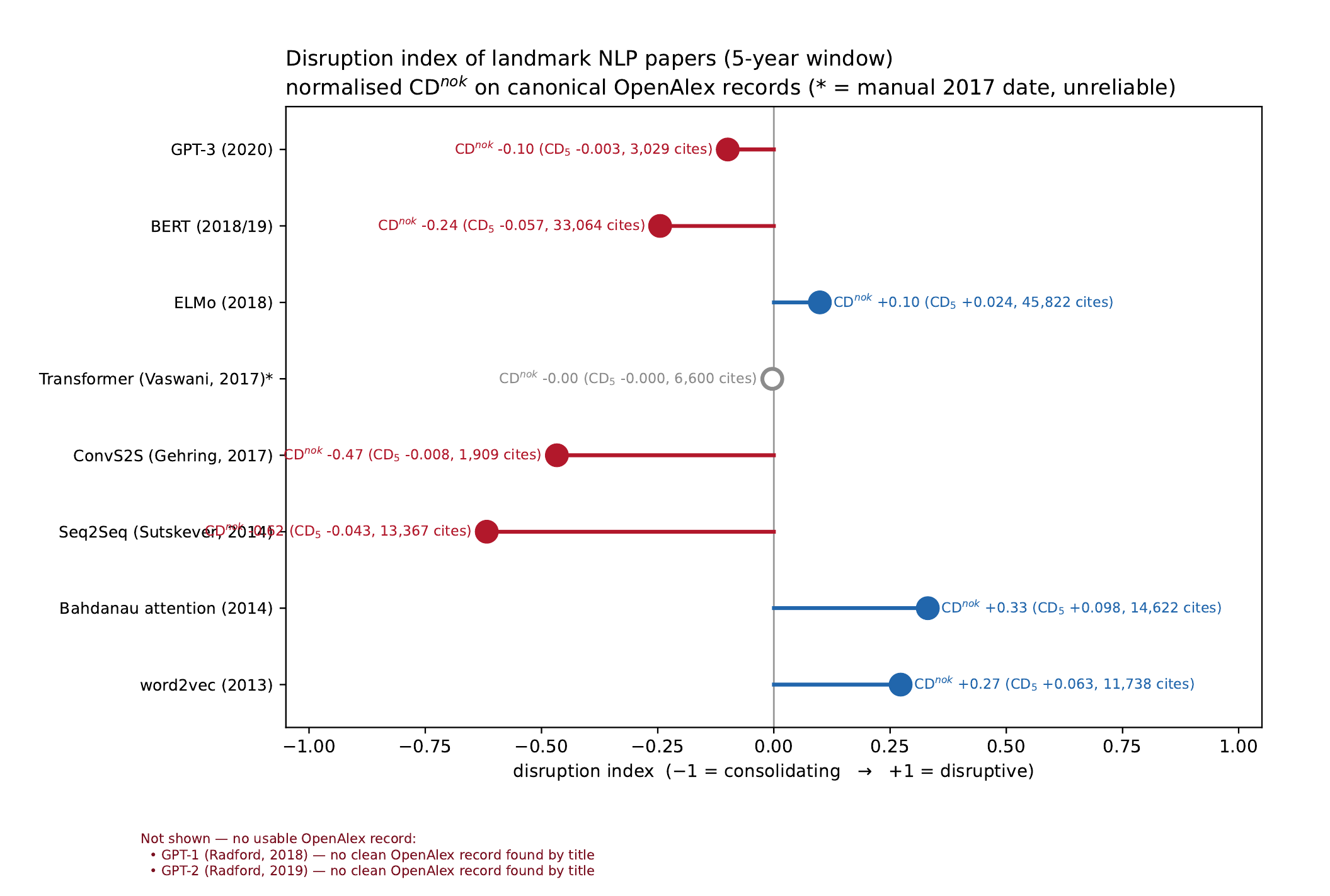}{0.9}
\caption{Disruption index of landmark NLP papers (normalised $\CDnok$). Values are small and mixed (e.g.\ Bahdanau disruptive, BERT consolidating); ``Attention Is All You Need'' has no clean OpenAlex record.}
\label{fig:landmarks}
\end{figure}

Figure~\ref{fig:landmarks} reports $\CD$ / normalised $\CDnok$ for individual landmark papers, supporting the P1b claim that the disruption index is small and mixed even for field-defining works, and that citation-record fragmentation (e.g.\ for the Transformer paper) further limits a naive citation-disruption monitor.

\section{Vocabulary-matched control}\label{app:control}

The equal-latency AUCs (Section~\ref{sec:bakeoff}) use a vocabulary-matched control: cs.CL papers that contain the challenger vocabulary (self-attention / attention / transformer) but lie outside the sequence-modelling shift-core ($n=1{,}356$). By construction the control is at least as saturated with challenger terms as the shift-core (challenger-term prevalence 1.00 vs 0.26), so a broad keyword counter cannot separate the two by vocabulary---indeed it collapses to AUC $=0.00$. The control is topically within cs.CL (embedding-centroid cosine to the shift-core 0.96, comparable to a random cs.CL sample at 0.97), i.e.\ not an easy off-topic foil. That the belief signal still separates the shift-core from this term-matched, topically-close control (AUC $\approx0.99$) is therefore evidence of content grounding, not vocabulary or topic leakage. One caveat, controlled for in Section~\ref{sec:bakeoff}: the strict term ``transformer'' occurs in 0.15 of the core and 0.00 of the control by cohort construction, so we do not treat the strict counter as a discriminator, and we additionally report the belief signal on the same top-cited focal set as the citation-based successors.

\section{Baseline implementations}\label{app:baselines}

All equal-latency baselines use the \emph{same} per-quarter grid, onset rule, and shift-vs-control AUC harness as the belief signal; none is tuned on held-out data. Where a released checkpoint was unavailable we used a faithful reimplementation and say so.
\begin{itemize}[leftmargin=*]
\item \textbf{N1 --- term prevalence.} Per-quarter share of papers whose title+abstract match the challenger regex (\texttt{self-attention|attention|transformer}); ``N1s'' restricts to the strict term \texttt{transformer}. No parameters.
\item \textbf{B1 --- rolling lexical surprise.} A bigram add-$k$ ($k=1$) language model estimated on the \emph{prior} quarters; a paper's surprise is its mean token negative-log-probability; the signal is the per-quarter mean. Vocabulary is refit each quarter on the trailing history (no future leakage).
\item \textbf{B5 --- embedding surprise.} Cosine distance of a paper's \texttt{text-embedding-3-small} (256-d) vector to the centroid of all field papers in the preceding four quarters; per-quarter mean.
\item \textbf{B2 --- reference atypicality} \citep{uzzi2013atypical}. Venue-pair $z$-scores of a paper's references against a degree-preserving within-year null (10 shuffles); paper score is the 10th-percentile ($z$) tail. Computed for cs.CV (its cs.CL reference pull is a separate job).
\item \textbf{B6 --- SPS-BERT} \citep{he2024predicting}. Faithful reimplementation of a learned title+abstract$\to$CD predictor: the cached sentence embedding (256-d) feeds a ridge head ($\alpha=10$) fit on $\CD$. \emph{Leakage discipline:} the head is trained \emph{only} on the 442 focal papers dated before the belief onset (pre-2016), whose $\CD$ five-year window is complete; it never sees post-onset papers. The signal is the predicted CD, aggregated per quarter.
\item \textbf{B7 --- embedding-disruptiveness} \citep{kim2026uncovering}. Because CD was computed from facet counts, the edges were materialized specially: control papers matched to OpenAlex, up to 100 citers pulled per focal paper, and 88k neighbour works embedded (256-d). The estimator is an embedding analogue of CD,
\[
\mathrm{EDM}(d)=\operatorname{mean}_{c\in\mathrm{citers}}\bigl[\cos(e_c,e_d)-\cos(e_c,\bar r_d)\bigr],
\]
with $\bar r_d$ the reference centroid; per-quarter mean over the focal set.
\end{itemize}
Two are supervised on or derived from CD (B6, B7), so they inherit CD's flatness by construction; the remainder (N1, B1, B5, B2) are CD-independent. None onsets for the shift (Table~\ref{tab:bakeoff}).

\end{document}